# MIRAS OR SRA'S – THE TRANSIENT TYPE VARIABLES

V.I. Marsakova[1], I.L. Andronov[2]

[1] Odessa National University named after I.I.Mechnikov, Odessa, Ukraine
*vmarsakova @ mail.ru*

[2] Department "High and Applied Mathematics", Odessa National Maritime University, Odessa, Ukraine
*tt_ari @ ukr.net*

**Abstract.** The group of Mira-type and semi-regular variables with similar periodicity (multiperiodicty) is analyzed. They have periods of 230–260 days and 140–150 days and show intervals of periodical (Mira-type) variability with relatively high amplitude and "semi-regular" (SR-type) small-amplitude oscillations. Results of periodogram analysis are represented.

Several variables of M and SR types drew our attention due to similar properties of their photometric behavior: intervals of periodical (Mira-type) variability with relatively high amplitudes turns to "semi-regular" (SR-type) small-amplitude oscillations with not so prominent periodicity on their light curves. Some star s show secular amplitude decreasing (such as Y Per and V Boo at the Fig. 1) that is typically for semi-regular or transient type variables [1].

For some of these variables, very similar periodicity (or multi-periodicity) was found.

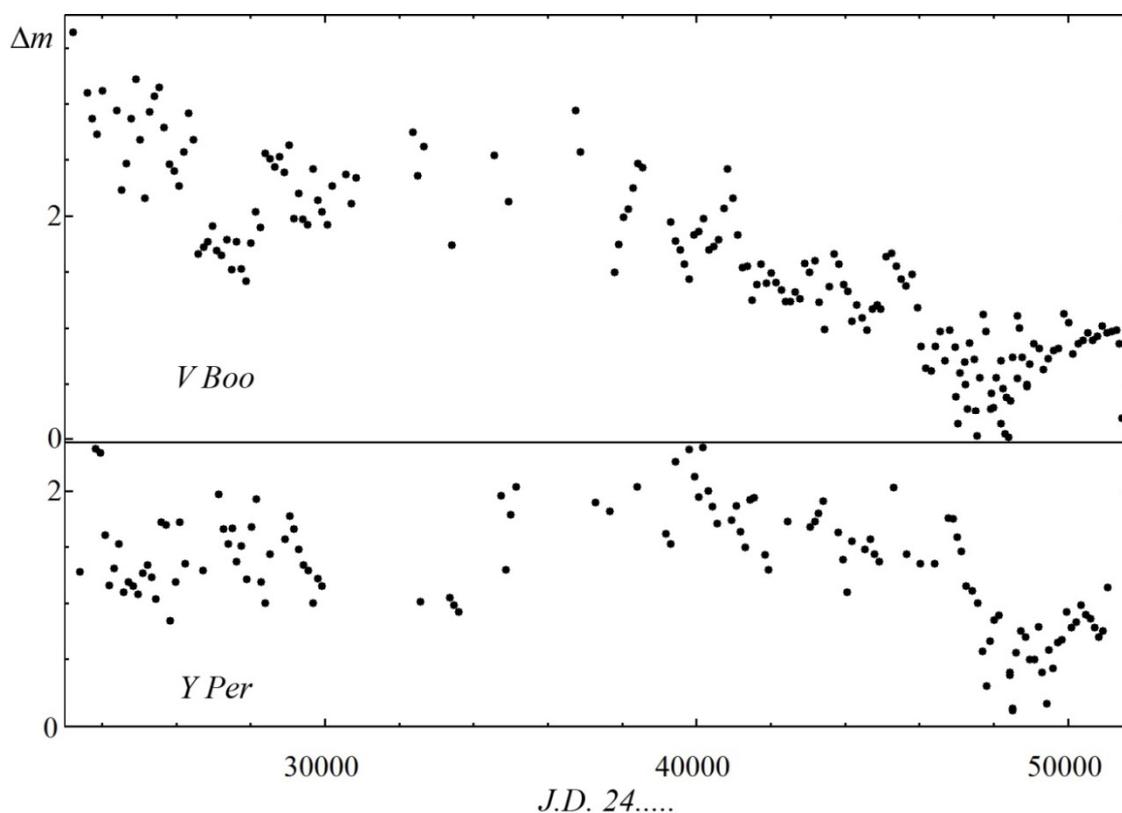

Fig. 1 Amplitude changes of V Boo and Y Per. Light curve for Y Per is shown in [2]. The amplitude is measured twice in cycle at ascending and descending branches.

Four variables were carefully analyzed [3], [4]. The methodic [5] have been used. Such we have studied their by using different methods of time-series analysis, such as the periodogram analysis using trigonometric polynomial fit [6], wavelet analysis and individual cycle characteristics analysis using the running parabola fit [7]. We have used the amateur observations from AAVSO database.

For other stars we give the results of preliminary periodogram analysis in the Table 1.

Several period values in the column $P_1$, may be caused by period changes during 100-years interval typical for long periodic variables. The values of test-function S are shown in the brackets. For peaks with similar height, several period ratios are given. S Aql shows an inversion of period sequence, but period ratio also is calculated as bigger-to-smaller.

Table 1. Results of periodogram analysis.

| Variable | $P_1$, (S) | $P_2$, (S) | Period ratio | GCVS classification [8] | Spectral class [8] |
|---|---|---|---|---|---|
| T Col | 226.1 (0.60) 229.8 (0.11) | 139.6 (0.06) | 1.62 | M | M3e-M6e |
| DN Her | 225.3 (0.52) 230.7 (0.18) | 139.3 (0.13) | 1.62 | M | |
| EL Lyr | 235.8 (0.57) 230.8 (0.31) | 143.3 (0.20) | 1.65 1.61 | M | |
| S Tri | 249.5 (0.16) 240.4 (0.08) | 148.1 (0.07) | 1.68 1.62 | M | M2e |
| Y Per | 253 (0.32). 245.3 (0.08) | 149.4 (0.04) | 1.70 | M | C4,3e |
| S Sex | 254.7 (0.32) 258.8 (0.26) 264.6 (0.20) | 150 (0.15) | 1.70 1.72 1.76 | M | M2-M5e |
| UZ Hya | 266.3 (0.59) | 153.8 (0.2) | 1.73 | M | M4e |
| AN Peg | 272.0 (0.6) | 156.0 (0.26) | 1.74 | M | M5 |
| S Aql | 146.7(0.48) | 245.2 (0.18) | 1.67 | SRa | M3-M5.5e |
| RU And | 234.3 (0.08) 245.7 (0.07) | 124,7 (0,03) 146,8 (0,02) | 1.87 1.67 | SRa | M5-M6e |
| V Boo | 257.5(0.49) 260 (0.17) | 151 (0.07) | 1.70 | SRa | M6e |
| X Mon | 257.6 (0.37) 260.8 (0.07) | 151.1 (0.06) | 1.71 | SRa | M1e-M6ep |
| RR Her | 236.7 (0.24) 261.3 (0.05) | 143.5 (0.05) | 1.61 | SRb | C5,7e-8,1e |
| ST Her | 256.5 (0.07) | 150.8 (0.06) 152.1 (0.06) | 1.70 1.68 | SRb | M6-M7 |
| S Sct | 268.6 (0.03) | 151.3 (0.02) | 1.77 | SRb | C6,4 |

Typical periodograms are shown at the Fig. 2-3 for the Mira-type variable EL Lyr and semi-regular ST Her.

As the result, very similar multi-periodicities were detected: the periods 230–260 days are represented by all these stars as well as the periods 140–150 days. Period ratio 1,7 is typical for some long-periodic stars, but a bigger ratios also present [9]. So among SRa and M-variables they may form the separate group of "transient" variables (notable that all these semi-regular stars show intervals of "Mira-like" variability).

But there are some questions for further studies: Is this similar periodicity an evidence of close evolutionary stages? Is this stage long-lasting, if we observe many variables with these periods?

This study is a part of the projects "Inter-Longitude Astronomy" [10] and "Ukrainian Virtual Observatory" [11].

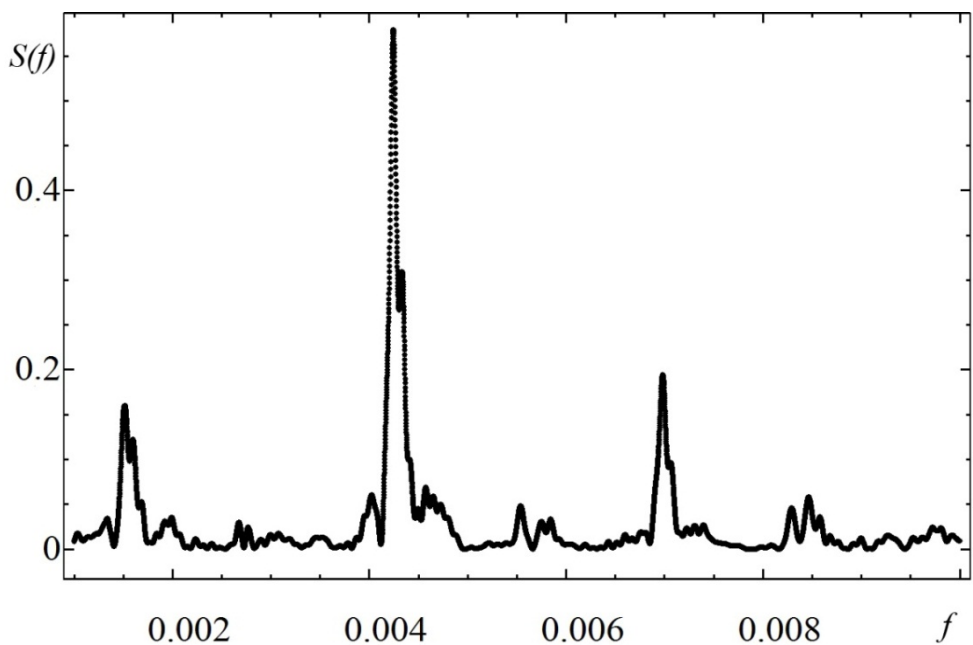

Fig. 2. Periodogram for EL Lyr.

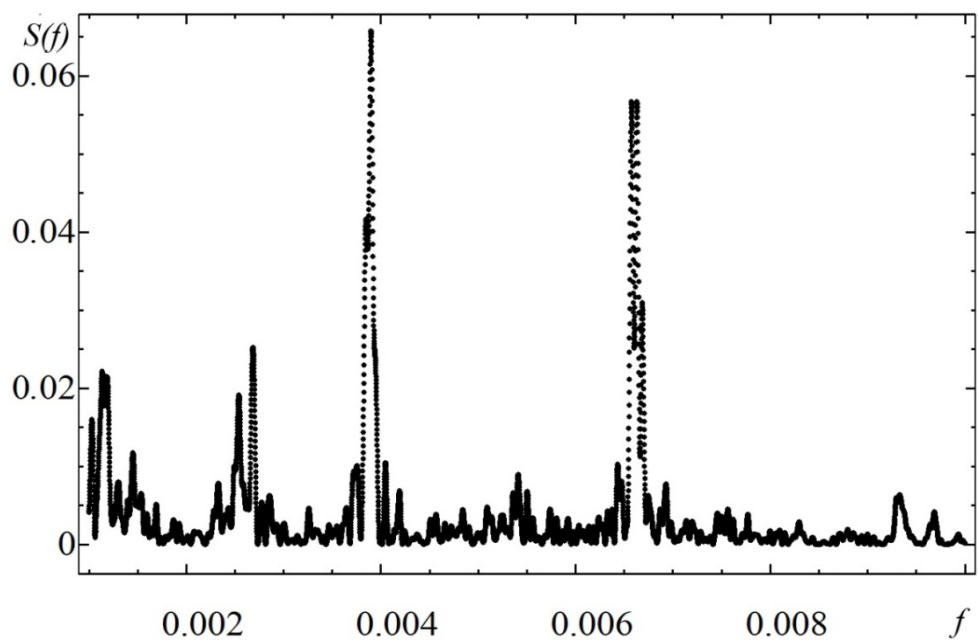

Fig. 3. Periodogram for ST Her.